\documentclass[superscriptaddress,secnumarabic,eqsecnum,amssymb,amsmath,nobibnotes,aps,prd,showkeys,showpacs,nofootinbib,preprint]{revtex4}%

\usepackage{graphics}
\usepackage{latexsym}
\usepackage{amsmath}
\usepackage{amssymb}
\usepackage{eufrak}
\usepackage{euscript}
\usepackage{pstricks}
\usepackage{graphics}
\usepackage{graphicx}
\usepackage{picture}

\newcommand{\bee}{\begin{equation}}
\newcommand{\eee}{\end{equation}}
\newcommand{\baa}{\begin{eqnarray}}
\newcommand{\eaa}{\end{eqnarray}}
\def\ni{\noindent}
\def\no{\nonumber}

\begin{document}

\count255=\time\divide\count255 by 60 \xdef\hourmin{\number\count255}
  \multiply\count255 by-60\advance\count255 by\time
 \xdef\hourmin{\hourmin:\ifnum\count255<10 0\fi\the\count255}


\title{{\Large  Gauge invariant actions for the noncommutative phase-space relativistic particle}}

\author{Everton M. C. Abreu}\email{evertonabreu@ufrrj.br}
\affiliation{Grupo de F\' isica Te\'orica e Matem\'atica F\' isica, Universidade Federal Rural do Rio de Janeiro, \\ 23890-971, Serop\'edica, RJ, Brazil}
\affiliation{Departamento de F\' isica, Universidade Federal de Juiz de Fora, 36036-330, Juiz de Fora, MG, Brazil}

\author {Cresus F. L. Godinho}\email{crgodinho@ufrrj.br}
\affiliation{Grupo de F\' isica Te\'orica e Matem\'atica F\' isica, Universidade Federal Rural do Rio de Janeiro, \\ 23890-971, Serop\'edica, RJ, Brazil}

\vskip 0.55in

\vskip 0.2in

\vskip .1in
\date{\today}
\vskip .95in


\begin{abstract}
\noindent  Our main interest here is to analyze the gauge invariance issue concerning the noncommutative relativistic particle.  Since the analysis of the constraint set from Dirac's point of view classifies it as a second-class system, it is not a gauge theory.  Hence, 
the objective here is to obtain gauge invariant actions linked to the original one.  However, we have two starting points, meaning that firstly we will begin directly from the original action and, using the Noether procedure, we have obtained a specific dual (gauge invariant) action.     Following another path, we will act toward the constraints so that  we have carried out the conversion of second to first-class constraints through the Batalin-Fradkin-Fradkina-Tyutin formalism, obtaining the second gauge invariant Lagrangian.
\end{abstract}


\newpage
\setcounter{page}{1}

\maketitle

\section{Introduction}\label{sec:intro}

To find a way that allows us to be free of the infinities that dwell in quantum field theory (QFT) is one of the great challenges in theoretical physics today.  Another challenge is to unify the ideas of general relativity  and the ones of the quantum mechanics, where it is believed to be the physics of the early Universe physics.  One of the paths that was believed to be the way to succeed in both problems is to work in a space-time where the multiplication of two position coordinates operators is not commutative.  In other words, where the commutation between two coordinates operators is not zero.  Of course we are talking about noncommutative  (NC) space-times (or phase spaces) in $N$-dimensional formulations \cite{reviews}.  There are also several NC approaches where we have a nonzero positions and/or momenta coordinates.  And where the NC parameter is constant or it makes part of the space-time or phase-space,  being a coordinate operator of an extended Hilbert space.  

The first known published paper that explored a NC space-time was carried out by Snyder \cite{snyder} in order to free QFT from infinities.   Short time later, Yang \cite{yang} demonstrated that the infinities were still there and this fact put Snyder's published ideas to sleep for more than forty years.  

At the end of the last century, motivated by the results brought by string theory, where the authors \cite{sw} found that the resulting algebra is a NC one (reproduced through symplectic formalism in \cite{bg}) the NC ideas found that the phenomenon of noncommutativity in space-times appears in many physical systems.   Abelian and non-Abelian theories were found to be affected by NC features as can be checked in recent reviews \cite{reviews}.  Even in IR/UV divergences we can encounter  contributions due to a NC space-time. 

Motivated by the possibility to understand the early Universe physics through a non-constant NC parameter \cite{faizal}, Doplicher, Fredenhagen and Roberts introduced the currently so-called DFR algebra \cite{dfr}.  In this formalism, the NC parameter is a coordinate operator in an extended Hilbert space.  Recently it was shown that this NC coordinate operator has an associated momentum \cite{nossos}. Some recent developments can be seen in \cite{nos}. 


Nowadays it is common knowledge that gauge invariance is a fundamental stone in standard model theory. Consequently, the investigation of how to obtain models that are gauge invariant is an important procedure in several areas of research in theoretical physics.

In some mechanical models, the NC geometric approach concerning the spatial variables can be obtained \cite{varios} from the resulting canonical quantization of the basic dynamical second-class constrained systems, in Dirac's approach.   After taking into account the constraints that appear in the model, the nontrivial brackets between the position coordinates can be calculated as the Dirac brackets.  A recurrent problem of the well known mechanical NC systems is the absence of relativistic invariance since the NC parameter is a constant matrix.

In this paper we review the Dirac's constraint analysis of a system where a NC space-time relativistic particle \cite{alexei0} is described in order to obtain gauge invariance.  In this system we have second class constraints, following Dirac's constraint nomenclature and we have used the BFFT \cite{bff,BT}  procedure to convert these second to first-class constraints. Although the position coordinates do not obey the DFR algebra, the phase-space carries a NC coordinate which is time dependent.  So, the model analyzed here and the DFR approach models share only the fact that the NC object is a coordinate of the phase-space.  Both characteristic algebras are completely different.

This paper is organized in the following manner.  In the next section, we have discussed briefly some features of NC classical mechanics.
After that, in section 3, we have explained the system discussed here, the NC relativistic particle.  In section 4, we have depicted the BFFT constraint conversion analysis.
In section 5 we have discussed the constraints of our system from the BFFT point of view.  In section 6, we have used the Noether ideas to obtain the first-class gauge invariant action.  In section 7, the BFFT method was used to obtain the second gauge invariant action.
The Conclusions and final remarks were described in Section 8.


\section{Noncommutative classical mechanics}

In the name of self-containment, in this section we will describe the main steps of the NC classical mechanics in order to explain the structure of the model explored here \cite{alexei0}.  The different thing about the description of this model is the fact, as we said before, that the NC object has coordinate properties and it is time dependent.  

Another relevant feature is the fact that, being a coordinate, it has an associated momentum, which brings a new feature that does not appear in other NC classical mechanics formulations.  These last ones consider in general, the NC object as an algebraic and constant parameter that only breaks the position coordinates commutation relation (and/or the momenta commutation relation) and, being a constant, has no dynamics.  Of course, classically speaking, we mean the commutation relations concerning the Poisson brackets relations.  The following description has these issues discussed.

\subsection{Noncommutative version of an arbitrary nondegenerate mechanics: a review}

Let us begin by analyzing a nondegenerate mechanical system with the configuration space variables $q^A(t), ~
A=1,2, \ldots , n$, and the Lagrangian action \cite{alexei0}
\begin{eqnarray}\label{1}
S=\int dt L(q^A, ~ \dot q^A)\,\,.
\end{eqnarray}

\ni Thanks to the nondegenerate feature of the system, there are no constraints
in the Hamiltonian formulation. In this way, let $p_A$ be the conjugated momentum for
$q^A$, and the Hamiltonian action can be written as
\begin{eqnarray}\label{2}
S_H=\int dt \left[ p_A\dot q^A-H_0(q^A, ~ p_A)\right]\,\,.
\end{eqnarray}

\ni The equations of motion that follow from Eq. (\ref{1}) and (\ref{2}) can be shown to be
equivalent.  Namely, they also remain equivalent for any degenerated system 
\cite{gt,alexei4}.   In this case, the Hamiltonian carries the Lagrangian
multipliers.   In the same way, we can describe the initial system (\ref{1}) through the first order Lagrangian action, which is given by
\begin{eqnarray}
\label{3}
S_1=\int dt \left[v_A\dot q^A-H_0(q^A, ~ v_A)\right] \,\,,
\end{eqnarray}

\ni where $q^A(t)$ and $v_A(t)$ are the configuration space variables of the
system.   

We can see clearly that the Lagrangians in Eqs. (\ref{1}), (\ref{3})
are equivalent. As a matter of fact, writing the conjugated momentum for the
variables $q^A$ and $v_A$ as $p_A$ and $\pi^A$, respectively,  we can find the Hamiltonian
expression for the action (\ref{3}), the second-class constraints
$$p_A-v_A=0,\quad \mbox{and}\quad \pi^A=0\,\,.$$   
After the introduction of the corresponding Dirac brackets,
we can deal with the constraints as strong equations. Then the
Hamiltonian formulation for (\ref{3}) is the same as for (\ref{1}),
i.e., Eq. (\ref{2}).

The NC version of the system (\ref{1}) can be depicted by the
following Lagrangian 
\begin{eqnarray}\label{4}
S_N=\int dt \left[v_A\dot q^A-H_0(q^A, ~ v_A)+
\dot v_A\theta^{AB}v_B \right]\,\,,
\end{eqnarray}

\ni where $\theta^{AB}$ is a constant matrix. It seems to be the NC parameter for the variables $q^A$.

We will now analyze the model (\ref{4}) in the Hamiltonian formulation (the interested reader can see 
\cite{alexei3} for details). All the equations that compute the momenta are the primary constraints of the model, namely,
$p_A$ and  $\pi^A$ are the conjugated momenta for the variables $q^A$ and  $v_A$, respectively,
\begin{eqnarray}\label{5}
G_A\equiv p_A-v_A=0, \qquad
T^A\equiv\pi^A-\theta^{AB}v_B=0\,\,,
\end{eqnarray}

\ni where the Poisson bracket algebra is a second-class one, as we can see from 
\begin{eqnarray}\label{6}
\{G_A, G_B\}=0\,\,, \qquad \{T^A, T^B\}=
-2\theta^{AB}\,\,, \qquad
\{G_A, T^B\}=-\delta_A^B\,\,.
\end{eqnarray}

The constraints can be considered as being a transition to Dirac's bracket. After that, one can take the variables $(q^A, ~ p_A)$ as the
physical one, while $(v_A, ~ \pi^A)$ can be omitted from the analysis by using Eq. (\ref{5}).  For example, the resulting NC system has the same number of physical degrees of freedom as the initial system $S$, i.e., $q^A$ and $p_A$. 
The equations of motion of the system are conserved, in other words, they are the same as for the initial
system $S$, modulo the term directly connected to the NC parameter
$\theta^{AB}$
\begin{align}
\label{7}
&\dot q^A=\frac{\partial H_0}{\partial p_A}-2\theta^{AB}
\frac{\partial H_0}{\partial q^B}\,\,, \nonumber \\
&\dot p_A=-\frac{\partial H_0}{\partial q^A}\,\,,
\end{align}

\ni where $H_0(q, ~ p)=H_0(q, ~ v)|_{v\rightarrow p}$, where we have used that the physical variables have the brackets
\begin{eqnarray}\label{8}
\{q^A, q^B\}=-2\theta^{AB}\,\,, \qquad
\{q^A, p_B\}=\delta^A_B\,\,,
\qquad \{p_A, p_B\}=0 \,\,.
\end{eqnarray}

\ni and we can see that the brackets of the configuration space coordinates are NC.  

To provide the quantization of the resulting system, one way is to construct variables which have the canonical brackets. For the case under consideration they can be written as
\begin{eqnarray}\label{9}
\tilde q^A=q^A-\theta^{AB}p_B \qquad \mbox{and} \qquad \tilde p_A=p_A\,\,,
\end{eqnarray}

\ni which obey $\{\tilde q, \tilde q\}=\{\tilde p, \tilde p\}=0, ~
\{\tilde q, \tilde p\}=1$ and the commutativity is recovered.  The above relations are well known as Bopp shift. 
The equations of motion in terms
of these modified commutative variables have the standard form
\begin{eqnarray}\label{10}
\dot{\tilde q}^A=\{\tilde q^A, \tilde H_0\}, \qquad
\dot{\tilde p}_A=\{\tilde p_A, \tilde H_0\}\,\,,
\end{eqnarray}

\ni where $\tilde H_0=H_0(\tilde q+\theta\tilde p, ~ \tilde p)$. It can lead us to formulations of
quantum mechanics with the Moyal-Weyl product (\cite{alexei2} and references therein) where the standard product is substituted by the Moyal-Weyl product, 
\begin{align}\label{11}
H_0(\tilde q^A+\theta^{AB}\tilde p_B, ~ \tilde p_B)\Psi(\tilde q^C)=
H_0(\tilde q^A, ~ \tilde p_B)*\Psi(\tilde q^C)\,\,,
\end{align}

\ni which is a so-called star-product, a well defined class of product which, among other properties, has the one that says that inside an integral, the Moyal-Weyl product is equal to the standard one.   However, notice that this substitution affects only the standard product.  After that, a well known map named as the Seiberg-Witten map, makes the connection between commutative and NC variables.

What was described so far can be also
used in some degenerated systems \cite{alexei0}. 
We know that a set of variables can enter into the initial Lagrangian without the time derivatives,
and, in this way, they can be considered as
the Lagrange multipliers of the Hamiltonian formulation.
Then the system admits the so-called first order  Lagrangian formulation given in
Eq. (\ref{3}). Well known examples are the relativistic
particle and the string theory \cite{alexei4}.   The procedure
can be used to analyze the spinning particle \cite{bm} and the superparticle \cite{cbs} models,
since both models are supersymmetric \cite{jj}.
If the relativistic invariance would be
introduced in the initial formulation, a small change concerning
the procedure is required in order to keep the symmetry in the NC version. More details can be found in \cite{alexei0}.

\section{Noncommutative relativistic particle}

In this section, we will describe the NC relativistic particle described in Deriglazov's paper \cite{alexei0}.
The configuration space defined by the variables $x^\mu(\tau), ~ v^\mu(\tau), ~ e(\tau)$, $\theta^{\mu\nu}(\tau)$ and the Lagrangian written as 
\begin{eqnarray}\label{12}
S=\int d\tau\left[\dot x^\mu v_\mu-\frac{e}{2}(v^2-m^2)+
\frac{1}{\theta^2}\dot v_\mu\theta^{\mu\nu}v_\nu\right]\,\,,
\end{eqnarray}

\ni are the basics ingredients of our analysis.   We will use that $\theta^2\equiv\theta^{\mu\nu}\theta_{\mu\nu}$ and $\eta^{\mu\nu}=
(+,-, \ldots ,-)$. To include the term $\theta^2$
in the denominator of the last term in (\ref{12}) has the same effect as the einbein in the massless particle action, i.e., 
$$L=\frac{1}{2e}\dot x^2\,\,.$$ 

\ni Namely, it excludes the degenerated gauge $e=0$ and the action is clearly manifestly invariant under Poincare transformations
\begin{eqnarray}\label{13}
x'^\mu=\Lambda^\mu{}_\nu x^\nu+a^\mu, \quad v'^\mu=\Lambda^\mu{}_\nu
v^\nu \quad \mbox{and} \quad
\theta '^{\mu\nu}=\Lambda^\mu{}_\rho\Lambda^\nu{}_\sigma
\theta^{\rho\sigma} \,\,.
\end{eqnarray}

\ni It is easy to check that the local symmetries of the model are reparametrizations where
$\theta^{\mu\nu}$ is deemed as the scalar variable.  
We also have the following transformations with the parameter $\epsilon_{\mu\nu}(\tau)=-\,\epsilon_{\nu\mu}(\tau)$
\begin{eqnarray}\label{14}
\delta x^\mu=-\epsilon^{\mu\nu}v_\nu \quad \mbox{and} \quad
\delta\theta_{\mu\nu}=-\theta^2\epsilon_{\mu\nu}+
2\theta_{\mu\nu}(\theta\epsilon)\,\,.
\end{eqnarray}

\ni  In order to analyze the sector which carries the physical information of this constrained system, we can rewrite it with a Hamiltonian form.
Let us begin with the action (\ref{12}), where we can find, in the Hamiltonian framework, the primary constraints
\begin{align}\label{15}
&G^\mu\equiv p^\mu-v^\mu=0\,\,, \qquad
T^\mu\equiv\pi^\mu-\frac{1}{\theta^2}\theta^{\mu\nu}v_\nu=0\,\,, \cr
&p_{\theta}^{\mu\nu}=0\,\,, \qquad \qquad \qquad\:\: p_e=0\,\,
\end{align}

\ni and the Hamiltonian is given by
\begin{eqnarray}\label{16}
H=\frac{e}{2}(v^2-m^2)+\lambda_{1\mu}G^\mu+\lambda_{2\mu}T^\mu+
\lambda_ep_e+\lambda_{\theta\mu\nu}p_{\theta}^{\mu\nu}\,\,.
\end{eqnarray}

\ni It is important to notice that $p$ and $\pi$ are the conjugated momenta for $x$ and $v$, respectively.   And $\lambda$ are the
Lagrangian multipliers for the constraints. The next step is to compute the secondary constraint
\begin{eqnarray}\label{17}
v^2-m^2=0\,\,.
\end{eqnarray}

\ni The equations of motion for calculating the Lagrangian multipliers are
\begin{align}\label{18}
&\lambda_2^\mu=0, \cr
&\lambda_1^\mu=ev^\mu+\frac{2}{\theta^2}(\lambda_\theta v)^\mu-
\frac{4}{\theta^4}(\theta\lambda_{\theta})(\theta v)^\mu\,\,.
\end{align}

\ni In our case we have not the so-called tertiary constraints.  Namely, the consistency conditions for the secondary constraints gives no new constraint.  The equations of motion follow from (\ref{16})-(\ref{18}).  
Considering the variables $x$ and $p$ we have that
\begin{align}\label{19}
&\dot x^\mu=ep^\mu+\frac{2}{\theta^2}(\lambda_\theta v)^\mu-
\frac{4}{\theta^4}(\theta\lambda_{\theta})(\theta v)^\mu, \cr
&\dot p^\mu=0 \,\,.
\end{align}

\ni Finally, the Poisson brackets concerning the constraints are given by
\begin{align}
\label{20}
&\{G^\mu, G^\nu\}=0\,\,, \qquad \qquad \{T^\mu, T^\nu\}=
-\frac{2}{\theta^2}\theta^{\mu\nu}\,\,, \nonumber \\
&\{G_\mu, T^\nu\}=-\delta_\mu^\nu\,\,, \qquad
\{T_\mu, p_\theta^{\rho\sigma}\}=-\frac{1}{\theta^2}
\delta_\mu^{[\rho}v^{\sigma ]}+
\frac{4}{\theta^4}(\theta v)_\mu\theta^{\rho\sigma}\,\,.
\end{align}

\ni We will consider the set of constraints $G^\mu$ and $T^\mu$ as a second-class subsystem which can
be considered into account by the transition to the Dirac bracket. Hence, the
remaining constraints can be classified following their properties according to the Dirac brackets. 
The consistency of the procedure is confirmed by well known theorems \cite{gitman}.  The corresponding Dirac brackets for the constraint algebra \eqref{20} can be found in \cite{alexei0}.

In the following sections we will analyze the gauge invariance issue through the point of view of two different approaches.  The conversion of all the second-class constraints to first-class one, bringing to the surface a gauge invariant action.  The first one is through the second-class Lagrangian itself.  We will impose a simple gauge transformation and, after an iterative methodology, the Noether formalism, we will construct a gauge invariant action.   The analysis of the constraints will be presented too.  After that, we will make a second gauge invariance investigation of the second-class constraints conversion to first-class one.  The final Lagrangian is gauge invariant, of course.  A comparison between both final actions will be accomplished.


\section{BFFT brief review}

Let us consider a system described by a Hamiltonian $H_0$ in a
phase-space $(q^i,p^i)$ with $i=1,\dots,N$. Here we suppose that the
coordinates are bosonic.  It can be shown that extensions to include fermionic degrees of
freedom and to the continuous case can be done in a straightforward
way. It is also supposed that there just exist second-class
constraints. Denoting them by $T_a$, with $a=1,\dots ,M<2N$, we have

\begin{equation}
\bigl\{T_a,\,T_b\bigr\}=\Delta_{ab}\,\,,
\label{2.1}
\end{equation}

\noindent where $\det(\Delta_{ab})\not=0$. 

As was said, the general purpose of the BFFT formalism is to convert
second-class constraints into first-class ones. This is achieved by
introducing canonical variables, one for each second-class constraint.
The connection between the number of second-class constraints and
the new variables in a one-to-one correlation is to keep the same
number of the physical degrees of freedom in the resulting extended
theory. We denote these auxiliary variables by $\eta^a$ and assume
that they have the following general structure

\begin{equation}
\bigl\{\eta^a,\,\eta^b\bigr\}=\omega^{ab}\,\,,
\label{2.2}
\end{equation}

\noindent where $\omega^{ab}$ is a constant quantity with
$\det\,(\omega^{ab})\not=0$. The obtainment of $\omega^{ab}$ is
embodied in the calculation of the resulting first-class constraints
that we denote by $\tilde T_a$. Of course, these depend on the new
variables $\eta^a$, namely

\begin{equation}
\tilde T_a=\tilde T_a(q,p;\eta)
\label{2.3}
\end{equation}

\noindent and it is considered to satisfy the boundary condition

\begin{equation}
\tilde T_a(q,p;0)=\tilde T_a(q,p)\,\,.
\label{2.4}
\end{equation}

\noindent The characteristic of these new constraints in the BFFT method, as it
was originally formulated, is that they are assumed to be strongly
involutive, i.e.,

\begin{equation}
\bigl\{\tilde T_a,\,\tilde T_b\bigr\}=0\,\,.
\label{2.5}
\end{equation}

\noindent The solution of Eq.~(\ref{2.5}) can be achieved by considering
$\tilde T_a$ expanded as

\begin{equation}
\tilde T_a=\sum_{n=0}^\infty T_a^{(n)}\,\,,
\label{2.6}
\end{equation}

\noindent where $T_a^{(n)}$ is a term of order $n$ in $\eta$. Compatibility
with the boundary condition~(\ref{2.4}) requires 

\begin{equation}
T_a^{(0)}=T_a\,\,.
\label{2.7}
\end{equation}

The replacement of Eq.~(\ref{2.6}) into~(\ref{2.5}) leads to a set of
equations, one for each coefficient of $\eta^n$. We can list some of them as
\begin{eqnarray}
&&\bigl\{T_a,T_b\bigr\}
+\bigl\{T_a^{(1)},T_b^{(1)}\bigr\}_{(\eta)}=0
\label{2.8}\\
&&\bigl\{T_a,T_b^{(1)}\bigr\}+\bigl\{T_a^{(1)},T_b\bigr\}
+\bigl\{T_a^{(1)},T_b^{(2)}\bigr\}_{(\eta)}
+\bigl\{T_a^{(2)},T_b^{(1)}\bigr\}_{(\eta)}=0
\label{2.9}\\
&&\bigl\{T_a,T_b^{(2)}\bigr\}
+\bigl\{T_a^{(1)},T_b^{(1)}\bigr\}_{(q,p)}
+\bigl\{T_a^{(2)},T_b\bigr\}
+\bigl\{T_a^{(1)},T_b^{(3)}\bigr\}_{(\eta)}
\nonumber\\
&&\phantom{\bigl\{T_a^{(0)},T_b^{(2)}\bigr\}_{(q,p)}}
+\bigl\{T_a^{(2)},T_b^{(2)}\bigr\}_{(\eta)}
+\bigl\{T_a^{(3)},T_b^{(1)}\bigr\}_{(\eta)}=0
\label{2.10}\\
&&\phantom{\bigl\{T_a^{(0)},T_b^{(2)}\bigr\}_{(q,p)}+}
\vdots
\nonumber
\end{eqnarray}

\noindent  The notation $\{,\}_{(q,p)}$ and $\{,\}_{(\eta)}$, represents the
parts of the Poisson bracket $\{,\}$ relative to the variables
$(q,p)$ and $(\eta)$, respectively.

Equations above are used iteratively in the obtainment of the
corrections $T^{(n)}$ ($n\geq1$). Equation~(\ref{2.8}) shall give
$T^{(1)}$. With this result and Eq.~(\ref{2.9}), one calculates
$T^{(2)}$, and so on. Since $T^{(1)}$ is linear in $\eta$ we may
write
\begin{equation}
T_a^{(1)}=X_{ab}(q,p)\,\eta^b\,\,.
\label{2.11}
\end{equation}

\noindent Introducing this expression into Eq.~(\ref{2.8}) and using
Eqs.~(\ref{2.1}) and (\ref{2.2}), we can write that
\begin{equation}
\Delta_{ab}+X_{ac}\,\omega^{cd}\,X_{bd}=0\,\,.
\label{2.12}
\end{equation}

\noindent We notice that this equation does not give $X_{ab}$ univocally,
because it also contains the still unknown $\omega^{ab}$. What we
usually do is to choose $\omega^{ab}$ in such a way that the new
variables are unconstrained. It might be opportune to mention that
sometimes it is not possible to make a choice like that \cite{Barc2}.
In this case, the new variables are constrained. In consequence, the
consistency of the method requires an introduction of other new
variables in order to transform these constraints also into
first-class. This may lead to an endless process. However, it is
important to emphasize that $\omega^{ab}$ can be fixed anyway.

However, even one fixes $\omega^{ab}$ it is still not possible to
obtain a univocally solution for $X_{ab}$. Let us check this point.
Since we are only considering  bosonic coordinates\footnote{The
problem also exists for the fermionic sector.}, 
$\Delta_{ab}$ and $\omega^{ab}$ are antisymmetric quantities. So,
expression (\ref{2.12}) compactly represents $M(M-1)/2$ independent
equations.  On the other hand, there is no prior symmetry involving
$X_{ab}$ and they consequently represent a set of $M^2$ independent
quantities.

In the case where $X_{ab}$ does not depend on ($q,p$), it is easily
seen that $T_a+\tilde T_a^{(1)}$ is already strongly involutive for
any choice we make and we succeed obtaining $\tilde T_a$. If this
is not so, the usual procedure is to introduce $T_a^{(1)}$ into Eq.
(\ref{2.9}) to calculate $T_a^{(2)}$ and so on. At this point resides
a problem that has been the origin of some developments of the
method, including the adoption of a non-Abelian constraint algebra.
This occurs because we do not know {\it a priori} what is the best
choice we can make to go from one step to another. Sometimes it is
possible to figure out a convenient choice for $X_{ab}$ in order to
obtain a first-class (Abelian) constraint algebra in the first stage
of the process \cite{Banerjee3,Banerjee4}. It is opportune to mention
that in the work of reference \cite{Banerjee1}, the use of a
non-Abelian algebra was in fact a way of avoiding to consider higher
order of the iterative method. More recently, the method has been
used (in its Abelian version) beyond the first correction
\cite{Banerjee2} and we mention that sometimes there are problems in
doing this \cite{Barc1}.

Another point of the usual BFFT formalism is that any dynamic function
$A(q,p)$ (for instance, the Hamiltonian) has also to be properly
modified in order to be strongly involutive with the first-class
constraints $\tilde T_a$. Denoting the modified quantity by $\tilde
A(q,p;\eta)$, we then have
\begin{equation}
\bigl\{\tilde T_a,\,\tilde A\bigr\}=0\,\,.
\label{2.13}
\end{equation}

\noindent In addition, $\tilde A$ has also to satisfy  the boundary condition
\begin{equation}
\tilde A(q,p;0)=A(q,p)\,\,.
\label{2.14}
\end{equation}

The obtainment of $\tilde A$ is similar to what was done to obtain $\tilde T_a$, that is to say, we consider an expansion like
\begin{equation}
\tilde A=\sum_{n=0}^\infty A^{(n)}\,\,,
\label{2.15}
\end{equation}

\noindent where $A^{(n)}$ is also a term of order $n$ in $\eta$'s.
Consequently, compatibility with Eq.~(\ref{2.14}) requires that

\begin{equation}
A^{(0)}=A\,\,.
\label{2.16}
\end{equation}

\noindent The combination of Eqs.~(\ref{2.6}), (\ref{2.7}), (\ref{2.13}),
(\ref{2.15}), and (\ref{2.16}) gives the equations

\begin{eqnarray}
&&\bigl\{T_a,A\bigr\}
+\bigl\{T_a^{(1)},A^{(1)}\bigr\}_{(\eta)}=0
\label{2.17}\\
&&\bigl\{T_a,A^{(1)}\bigr\}+\bigl\{T_a^{(1)},A\bigr\}
+\bigl\{T_a^{(1)},A^{(2)}\bigr\}_{(\eta)}
+\bigl\{T_a^{(2)},A^{(1)}\bigr\}_{(\eta)}=0
\label{2.18}\\
&&\bigl\{T_a,A^{(2)}\bigr\}
+\bigl\{T_a^{(1)},A^{(1)}\bigr\}_{(q,p)}
+\bigl\{T_a^{(2)},\bigr\}
+\bigl\{T_a^{(1)},A^{(3)}\bigr\}_{(\eta)}
\nonumber\\
&&\phantom{\bigl\{T_a^{(0)},A^{(2)}\bigr\}_{(q,p)}}
+\bigl\{T_a^{(2)},A^{(2)}\bigr\}_{(\eta)}
+\bigl\{T_a^{(3)},A^{(1)}\bigr\}_{(\eta)}=0
\label{2.19}\\
&&\phantom{\bigl\{T_a^{(0)},A^{(2)}\bigr\}_{(q,p)}+}
\vdots
\nonumber
\end{eqnarray}

\noindent which correspond to the coefficients of the powers 0, 1, 2, etc. of
the variable $\eta$ respectively. It is just a matter of algebraic
work to show that the general expression for $A^{(n)}$ reads
\begin{equation}
A^{(n+1)}=-{1\over n+1}\,\eta^a\,\omega_{ab}\,X^{bc}\,G_c^{(n)}\,\,,
\label{2.20}
\end{equation}

\noindent  where $\omega_{ab}$ and $X^{ab}$ are the inverses of $\omega^{ab}$
and $X_{ab}$, and
\begin{equation}
G_a^{(n)}=\sum_{m=0}^n\bigl\{T_a^{(n-m)},\,A^{(m)}\bigr\}_{(q,p)}
+\sum_{m=0}^{n-2}\bigl\{T_a^{(n-m)},\,A^{(m+2)}\bigr\}_{(\eta)}
+\bigl\{T_a^{(n+1)},\,A^{(1)}\bigr\}_{(\eta)}\,\,.
\label{2.21}
\end{equation}

The general prescription of the usual BFFT method to obtain the
Hamiltonian is the direct use of relations (\ref{2.15}) and
(\ref{2.20}). This works well for system with linear constraints. For
nonlinear theories, where it may be necessary to consider all order
of the iterative process, this calculation might be quite
complicated.  There is an alternative procedure that drastically
simplifies the algebraic work. The basic idea is to
obtain the involutive forms for the initial fields $q$ and $p$
\cite{Banerjee5}. This can be directly achieved from the previous
analysis to obtain $\tilde A$. Denoting these by $\tilde q$ and
$\tilde p$ we have
\begin{equation}
H(q,p)\longrightarrow H(\tilde q,\tilde p)
=\tilde H(\tilde q,\tilde p)\,\,.
\label{2.22}
\end{equation}

\noindent It is obvious that the initial boundary condition in the BFFT
process, namely, the reduction of the involutive function to the
original function when the new fields are set to zero, remains
preserved. Incidentally we mention that in the cases with linear
constraints, the new variables $\tilde q$ and $\tilde p$ are just
shifted coordinates in the auxiliary coordinate $\eta$ \cite{Ricardo}.

Let us now finally consider the case where the first-class
constraints form an non-Abelian algebra, i.e.,
\begin{equation}
\bigl\{\tilde T_a,\,\tilde T_b\bigr\}=C_{ab}^c\,\tilde T_c\,\,.
\label{2.23}
\end{equation}

\noindent The quantities $C_{ab}^c$ are the structure constants of the
non-Abelian algebra. These constraints are considered to satisfy the
same previous conditions given by (\ref{2.3}), (\ref{2.4}),
(\ref{2.6}), and (\ref{2.7}). But now, instead of Eqs.
(\ref{2.8})-(\ref{2.10}), we obtain 

\begin{eqnarray}
C_{ab}^c\,T_c&=&\bigl\{T_a,T_b\bigr\}
+\bigl\{T_a^{(1)},T_b^{(1)}\bigr\}_{(\eta)}
\label{2.24}\\
C_{ab}^c\,T_c^{(1)}&=&\bigl\{T_a,T_b^{(1)}\bigr\}
+\bigl\{T_a^{(1)},T_b\bigr\}
\nonumber\\
&&+\,\bigl\{T_a^{(1)},T_b^{(2)}\bigr\}_{(\eta)}
+\bigl\{T_a^{(2)},T_b^{(1)}\bigr\}_{(\eta)}
\label{2.25}\\
C_{ab}^c\,T_c^{(2)}&=&\bigl\{T_a,T_b^{(2)}\bigr\}
+\bigl\{T_a^{(1)},T_b^{(1)}\bigr\}_{(q,p)}
\nonumber\\
&&+\bigl\{T_a^{(2)},T_b^{(0)}\bigr\}_{(q,p)}
+\bigl\{T_a^{(1)},T_b^{(3)}\bigr\}_{(\eta)}
\nonumber\\
&&+\bigl\{T_a^{(2)},T_b^{(2)}\bigr\}_{(\eta)}
+\bigl\{T_a^{(3)},T_b^{(1)}\bigr\}_{(\eta)}
\label{2.26}\\
&&\vdots
\nonumber
\end{eqnarray}

\noindent  The use of these equations is the same as before, i.e., they shall
work iteratively. Equation (\ref{2.24}) gives $T^{(1)}$.  With this
result and Eq. (\ref{2.25}) one calculates $T^{(2)}$, and so on. To
calculate the first correction, we assume it is given by the same
general expression (\ref{2.11}). Introducing it into (\ref{2.24}), we
now have
\begin{equation}
C_{ab}^c\,T_c=\Delta_{ab}+X_{ac}\,\omega^{cd}\,X_{bd}\,\,.
\label{2.27}
\end{equation}

\noindent  Of course, the same difficulties pointed out with respect to the
solutions of Eq.  (\ref{2.12}) also apply here, with the additional
problem of choosing the appropriate structure constants $C_{ab}^c$.

To obtain the embedding Hamiltonian $\tilde H(q,p,\eta)$ one cannot
use the simplified version discussed here for the Abelian case, embodied
into Eq. (\ref{2.22}), because the algebra is not strong involutive
anymore. We here start from the fact that the new Hamiltonian $\tilde
H$ and the new constraints $\tilde T$ satisfy the relation
\begin{equation}
\bigl\{\tilde T_a,\,\tilde H\bigr\}=B_a^b\,\tilde T_b
\label{2.28}
\end{equation}

\noindent where the coefficients $B_a^b$ are the same coefficients that may appear
in the consistency condition of the initial constraints, namely,
\begin{equation}
\bigl\{T_a,\, H\bigr\}=B_a^b\,T_b
\label{2.29}
\end{equation}

\noindent because in the limit of $\eta\rightarrow0$ both relations
(\ref{2.28}) and (\ref{2.29}) coincide. The involutive Hamiltonian
is considered to satisfy the same conditions
(\ref{2.14})-(\ref{2.16}). We then obtain that the general correction
$H^{(n)}$ is given by a relation similar to (\ref{2.20}), but now the
quantities $G_a^{(n)}$ are given by
\begin{eqnarray}
G_a^{(n)}&=&\sum_{m=0}^n\bigl\{T_a^{(n-m)},\,H^{(m)}\bigr\}_{(q,p)}
+\sum_{m=0}^{n-2}\bigl\{T_a^{(n-m)},\,A^{(m+2)}\bigr\}_{(\eta)}
\nonumber\\
&&+\,\,\bigl\{T_a^{(n+1)},\,A^{(1)}\bigr\}_{(\eta)}
-B_a^b\,T_c^{(n)}\,\,.
\label{2.30}
\end{eqnarray}

In the next section we will begin to analyze the constraints of our model based on the concepts of the BFFT method.  However, the computation of the gauge invariant Lagrangian will be accomplished after the next section.  The objective of this separation has pedagogical reasons since we deal here with two completely different methods to introduce gauge invariance.


\section{NC relativistic particle BFFT constraint analysis}

The described  NC phase-space is formed by \cite{alexei0}

\bee
\label{aaa}
\Sigma\,=\, \{x^{\mu}, v^{\mu},e, \theta^{\mu\nu}, p_{\mu}, \pi_{\mu}, p_{e}, p_{\mu\nu}^{\theta}\}
\eee

\ni and the second-class constraints are given by
\baa
\label{bbb}
G^{\mu}\,&=&\,p^{\mu}\,-\,v^{\mu}\,\approx \, 0\,\,, \no \\
T^{\mu}\,&=&\,\pi^{\mu}\,-\,\frac{\theta^{\mu\nu}}{\theta^2}\,v_{\nu}\,\approx \, 0\,\,, \no \\
V^{\mu\nu}\,&=&\,p_{\theta}^{\mu\nu}\,\approx \,0\,\,, \\
K_e\,&=&\,p_e\,\approx 0 \,\,, \no
\eaa

\ni where $G^{\mu}$ and $T^{\mu}$ are the second-class constraints.

The second BFFT step is to introduce the canonical variables, one for each second-class constraint.   In this way, we have a one-to-one relation between  the constraint and the new variable in order to not alter the number of the degrees of freedom in the final extended theory.  Let us denote these variables by $\omega^{\lambda}$ and
\baa
\label{ccc}
\tilde{G}^{\mu}\,&=&\,p^{\mu}\,-\,v^{\mu}\,-\,\omega^{\mu} \,\,, \no \\
\tilde{T}^{\mu}\,&=&\,\pi^{\mu}\,-\,\frac{{\theta}^{\mu\kappa}}{\theta^2}\,v_{\kappa}\,-\,p_{\omega}^{\mu}\,\,,
\eaa

\ni in order to have the following structure
\bee
\label{ddd}
\{\tilde{G}^{\mu}\,,\,\tilde{T}_{\nu}\}_P\,=\,\sum_{\rho} \bigglb(\frac{\partial \tilde{G}^{\mu}}{\partial q^{\rho}}\frac{\partial \tilde{T}_{\nu}}{\partial p_{\rho}}
\,-\,
\frac{\partial \tilde{G}^{\mu}}{\partial p_{\rho}}\frac{\partial \tilde{T}_{\nu}}{\partial q^{\rho}} \biggrb)\,\,,
\eee

\ni where
\baa
\label{eee}
q\,=\, \{x^{\mu}, v^{\mu},e, \theta^{\mu\nu}, \omega^{\mu}\} \,\,, \no \\
p\,=\, \{p_{\mu}, \pi_{\mu}, p_{e}, p_{\mu\nu}^{\theta},p^{\omega}_{\mu}\}\,\,.
\eaa

\ni  The new constraints, of course, depend on these new variables in order to
\baa
\label{fff}
\tilde{G}^{\mu}\,&=&\,\tilde{G}^{\mu}(\,q, p;\omega\,) \,\,, \no \\
\tilde{T}^{\mu}\,&=&\,\tilde{T}^{\mu}(\,q, p;p_{\omega}\,) \,\,,
\eaa

\ni and the reasonable boundary condition is
\baa
\label{ggg}
\tilde{G}^{\mu}(\,q, p;\omega\,=0)\,&=&\,\tilde{G}^{\mu}(\,q, p\,) \,\,, \no \\
\tilde{T}^{\mu}(\,q, p;p_{\omega}\,=0)\,&=&\,\tilde{T}^{\mu}(\,q, p\,) \,\,,
\eaa

\ni where we recover the old constraints and the algebra \eqref{20} is obeyed again.

From here we have set the stage to obtain new actions that are gauge invariant.  In the next section we will begin to attack this task computing this invariant action using the Noether procedure.  Since the NC parameter is not constant, which is different from the standard canonical noncommutativity literature, we will impose its gauge invariance.   Hence, it is not affected by the procedure, of course.
As we have said before, this method starts from the original action in Eq. \eqref{12}.  After that, we will use the ingredients of this very section to use the BFFT technique to work with the constraints to obtain a first-class new action.


\section{Gauge Invariance}

The action for the NC relativistic particle is described as a second-class system in Dirac's language \cite{alexei0}, and consequently it is not a gauge invariant model.  In this section we will use the Noether procedure \cite{iw} (a procedure used in the quantum interference of chiral actions \cite{solda}) to obtain a gauge invariant model derived from the NC relativistic particle one.

The initial Lagrangian, of course, is given by
\baa
\label{6.1}
{\cal L}_0\,&=&\, \dot{x}^\mu v_\mu \,-\, \frac e2 \Big(v^2 \,-\, m^2 \Big) \,+\, \frac{1}{\theta^2}\dot{v}_\mu \theta^{\mu\nu} v_\nu \no \\
&\Longrightarrow& \delta {\cal L}_0 \,=\, (\delta \dot{x}^\mu) v_\mu \,+\,\dot{x}^\mu (\delta v_\mu ) \,-\, \frac{\delta e}{2} \Big( v^2\,-\,m^2 \Big) \,-\,e v_\mu (\delta v^\mu) \no \\
\,&&\qquad\quad\,+\, \frac{1}{\theta^2} \Big[ (\delta \dot{v}_\mu v_\nu \,+\, \dot{v}_\mu ( \delta v_\nu ) \Big] \theta^{\mu\nu} \,\,,
\eaa

\ni where we have imposed firstly that $\delta \theta^{\mu\nu} (\tau) = 0$ and after that we will assume (following the procedure) the set of trivial gauge invariances  
\baa
\label{6.2}
\delta x^\mu = \delta v^\mu \,&=&\,\alpha^\mu (\tau) \no \\
\delta e &=& \beta (\tau) \,\,.
\eaa

In this way, after a few integral by parts calculations, we have that 
\bee
\label{6.3}
\delta {\cal L}_0 \,=\, J^\mu \alpha_\mu \,-\, J \beta\,\,,
\eee

\ni where $J^\mu$ and $J$ are the Noether currents
\baa
\label{6.4}
J^\mu &=& \dot{x}^\mu \,-\, \dot{v}^\mu \,-\, e v^{\mu} \,-\, 2 \dot{v}_\nu \bigglb( \frac{\theta^{\mu\nu}}{\theta^2} \biggrb)\,-\,v_\nu \bigglb( \frac{\theta^{\mu\nu}}{\theta^2} \biggrb)^\bullet\,\,,  \no \\
J &=& \frac 12 \Big( v^2 \,-\,m^2 \Big)\,\,,
\eaa

\ni where the dot in the last term of $J^\mu$ means time derivation of the expression inside the parenthesis.  Hence, obviously we have two Noether currents, which means that we have to introduce two auxiliary variables.

This is the next step of this iterative method, where new auxiliary variables are introduced in order to construct a new Lagrangian such that
\bee
\label{6.5}
{\cal L}_1 \,=\, {\cal L}_0 \,-\, J^\mu B_\mu \,+\, J\,C
\eee

\ni where $B_\mu$ is a space-time coordinate-like (will be eliminated) and $C$ is a scalar object. Both of them with be eliminated via equations of motion in due time. The variation of ${\cal L}_1$ is, trivially saying,
\bee
\label{6.6}
\delta {\cal L}_1 \,=\, \delta {\cal L}_0 \,-\, (\delta J^\mu ) B_\mu \,-\, J^\mu (\delta B_\mu ) \,+\, (\delta J )\, C \,+\, J (\delta C),
\eee

\ni and let us impose that the auxiliary variables variations are given by

\bee
\label{6.7}
\delta B_\mu \,=\,\alpha_\mu  \qquad \mbox{and} \qquad \delta C \,=\, \beta (\tau)
\eee

\ni and after that we can write that
\bee
\label{6.8}
\delta {\cal L}_1 \,=\,v^\mu \delta(B_\mu C) \,+\, \frac 12 e (\delta B^2_\mu ) \,-\, 2 (\delta \dot{B}_\nu ) B_\mu \frac{\theta^{\mu\nu}}{\theta^2} \,-\, (\delta B_\nu ) B_\mu \bigglb( \frac{\theta^{\mu\nu}}{\theta^2} \biggrb)^\bullet
\eee

\ni which is not zero, which means that the next iteration is
\baa
\label{6.9}
&&{\cal L}_2 \,=\, {\cal L}_1 \,-\, v^\mu B_\mu C \,-\, \frac 12 e B^2_\mu \,-\, 2 B_\mu \dot{B}_\nu \bigglb( \frac{\theta^{\mu\nu}}{\theta^2} \biggrb) \no \\
&&\Longrightarrow \delta {\cal L}_2 \,=\, - \frac 12 \delta (B^2_\mu C ) \,-\, 2 (\delta B_\mu ) \dot{B}_\nu \bigglb( \frac{\theta^{\mu\nu}}{\theta^2} \biggrb)
\eaa

\ni Hence,
\baa
\label{6.10}
&&{\cal L}_3 \,=\, {\cal L}_2 \,+\, \frac 12 B^2_\mu C \no \\
&& \Longrightarrow \delta {\cal L}_3 \,=\,-\,2\,\dot{B}_\nu (\delta B_\mu ) \bigglb( \frac{\theta^{\mu\nu}}{\theta^2} \biggrb)\,\,,
\eaa



\ni which means the end of the procedure since this final Lagrangian variation depends only on the auxiliary variables ($\theta$ is invariant).   However, we can see that $\delta {\cal L}_3$ is not zero.  Hence, we have to look for a solution that makes the Eq. \eqref{6.10} equal to zero.


This demand force us to search for a solution that fits into our gauge invariance necessity.   Since we have chosen the trivial gauge symmetry for the original variables, let us choose also a simple form for the auxiliary variable $B_\mu$ such that $\delta {\cal L}_3 =0$ and the gauge invariance was finally obtained.  So the simpler choice for $B_\mu$ is
\bee
\label{6.13}
B_\mu \,=\, \Big(B(\tau), 0, 0, 0 \Big)
\eee

\ni which guarantees that $\delta {\cal L}_3 =0$ since we can see that in Eq. \eqref{6.10} we must have two different components for $B_\mu$ but, only one of them is different from zero and for $\mu=\nu$ in \eqref{6.10} we have that $\theta^{\mu\nu}=0$.   Thus,
we have that the gauge invariance for the Lagrangian is obeyed and thus
\bee
\label{6.14}
{\cal L}_3 \,=\, {\cal L}_0 \,-\, J^\mu B_\mu \,+\, J C \,-\, v^\mu B_\mu C \,-\, \frac 12 e B^2_\mu \,+\, \frac 12 B^2_\mu C 
\,\,,
\eee

\ni since any product of different components of $B_\mu$ and $\theta^{\mu\nu}$ is zero using \eqref{6.13}, of course.   Notice that ${\cal L}_3  \rightarrow {\cal L}_0$ when we make $B_\mu=C=0$.  In this way we can say that $B_\mu$ and $C$ are unphysical variables.

Now we have to eliminate these auxiliary fields through their equations of motions, which are

\bee
\label{6.15}
\frac{\delta {\cal L}_3}{\delta B_\mu} \,=\, - J^\mu \,-\, v^\mu C \,-\,e B^\mu \,+\, B^\mu C
\,=\,0
\eee

\ni and
\bee
\label{6.16}
\frac{\delta {\cal L}_3}{\delta C} \,=\, J \,-\,v^\mu B_\mu \,+\, \frac 12 B^2_\mu \,=\,0
\eee

From \eqref{6.15} we have that the last term is zero since we have only $B_0$ and the other components of $B_\mu$ are zero.  So, from \eqref{6.15}
\bee
\label{6.17}
J^0 \,-\,v^0 C \,-\, e B^0 \,+\, B^0 C \,=\,0
\eee

\ni and from \eqref{6.16} we have that
\baa
\label{6.18}
&&B^2\,-\,2\,v^0\,B\,+\,2 J = 0 \no \\
&&\Longrightarrow B_{\pm}(\tau)\,=\, v^0 (\tau) \pm \sqrt{\Big(v^0(\tau)\Big)^2 \,-\,2 J}
\eaa

\ni and therefore we have that
\bee
\label{6.19}
B_\mu\,=\, \Big( B_\pm , 0,0,0 \Big) \,\,.
\eee

\ni Calculating $C$ from Eq. \eqref{6.16} we can write that
\baa
\label{6.20}
C_{\pm}\,&=& \frac{eB_\pm \,+\,J^0}{B_\pm \,-\,v^0} \no \\
&=&\,\pm\frac{e \Big[v^0 \pm \sqrt{(v^0)^2 \,-\, 2 J}\Big]\,+\,J^0}{\sqrt{(v^0)^2 \,-\,2 J}}\,\,,
\eaa

\ni where $J^0$ and $J$ were given in \eqref{6.4}.  Substituting Eqs. \eqref{6.4}, \eqref{6.19} and \eqref{6.20} into Eq. \eqref{6.14}, we have a gauge invariant action dual to the NC relativistic particle written in Eq. \eqref{6.1}.  In other words, we have a gauge invariant Lagrangian which is invariant under the imposed trivial gauge transformations given in \eqref{6.7}.  Remember that our objective is to construct a gauge invariant Lagrangian, which we succeed.

If, instead of making the choices written in \eqref{6.7}, we have chosen different gauge transformations like $\delta x^\mu = \alpha^\mu (\tau)$, $\delta v^\mu = \beta^\mu (\tau)$ and $\delta e = \omega(\tau)$ with three different auxiliary fields, an analogous calculation leads us to the condition that $\delta x^\mu = \delta v^\mu$ and so, the first trivial option is the correct one to construct a gauge invariant action using the Noether procedure.

So, the final gauge invariant action for the NC relativistic particle is given by 
\baa
\label{6.21}
{\cal L}_\pm \,&=&\,\dot{x}^\mu \,v_\mu \,-\, \frac e2 \Big(v^2 \,-\, m^2 \Big) \,+\, \frac{1}{\theta^2}\dot{v}_\mu \theta^{\mu\nu} v_\nu
\,-\, \Big( \dot{x}^0 - e v^0 - \dot{v}^0 - \frac{2}{\theta^2} \dot{v}_i \,\theta^{\,0i}\Big)B_\pm \no \\
&\,+\,& \frac 12 (v^2 - m^2 )\, C_\pm \,-\, v^0 B_\pm\, C_\pm \,-\, \frac 12 e B^2_\pm \,+\, \frac 12 B^2_\pm\, C_\pm \,\,,
\eaa

\ni where the $i$ means spatial component, $B=B_\pm =B_\pm(\tau)$ and $C_\pm=C_\pm(\tau)$ are given by Eqs. \eqref{6.18} and \eqref{6.20} respectively.  We can see that this Lagrangian is two fold, since we have two values for $B$ and $C$. 
We can see clearly from Eqs. \eqref{6.18} and \eqref{6.20} that as we have gained gauge invariance we have lost the explicit Lorentz covariance.


\section{The BFFT analysis}

Our extended phase space will be given by extra coordinates and momenta, respectively given by
\baa
\label{}
q&=& \{\,x^\mu,\, v^\mu,\, e,\, \theta^{\mu\nu},\, \omega^\mu ,\, \xi^\mu \,\} \,\,,\no \\
p&=& \{\, p^\mu,\, \pi^\mu,\, p_e ,\, p^{\theta}_{\mu\nu},\, p^{\omega}_\mu ,\, p^\xi_\mu \,\,,\}
\eaa

Following the BFFT procedure, the new variables are,

\baa
\label{7.2}
&&\tilde{x}^\mu  \longrightarrow x^\mu \,+\, (\omega_k \,+\, v_k ) \frac{\theta^{\mu k}}{\theta^2} \,-\,\pi^\mu\,\,, \nonumber \\
&&\tilde{\pi}^\mu  \longrightarrow \pi^\mu \,+\, (\omega_k \,+\, v_k ) \frac{\theta^{\mu k}}{\theta^2} \,+\,3 p^{\omega\mu}\,\,, \nonumber \\
&&\tilde{e} \longrightarrow e \,\,, \no \\
&&\tilde{v}^\mu \longrightarrow v^\mu  \\
&&\tilde{\theta}^{\mu\nu} \longrightarrow \theta^{\mu\nu}\,\,,  \no \\
&&\tilde{p}^\mu \longrightarrow p^\mu \,\,,  \no \\
&&\tilde{p}^{\theta\rho\sigma}  \longrightarrow p^{\theta\rho\sigma} \,+\, \bigglb[ \frac{(\delta^\rho_k\,W^{\sigma}\,-\,\delta^\sigma_k\,W^\rho )}{\theta^2} 
\,-\,4 \frac{\theta_{k\nu}}{\theta^4}W^\nu\,\theta^{\rho\sigma} \biggrb] \xi^k \,\,, \no 
\eaa

\ni where $W^\nu = \omega^\nu - v^\nu$ and we have that $$\{\tilde{a}, \tilde{b} \}^{PB}_{\phi=0}\,=\,\{ a, b \}^{DB}\,\,,$$ where $\phi$ represents the extra-fields, $\tilde{a}$ and $\tilde{b}$ are any variables in the extended phase space and $a,b$ are the standard variables.  We can see directly that when $\omega^\mu = p^{\omega}_{\mu} = 0$ we recover the old coordinates.
 We have from \cite{alexei0} that

\baa
\label{}
\{ \tilde{x},\, \tilde{x} \}_{\phi =0} &=& -\, 2\frac{\theta^{\mu\nu}}{\theta^2}\,\,, \no \\
\{ \tilde{x},\, \tilde{v}_nu \}_{\phi =0} &=& \delta^\mu_\nu\,\,, \no \\
\{ \tilde{x},\, \tilde{\pi}^\nu \}_{\phi =0} &=& -\,\frac{\theta^{\mu\nu}}{\theta^2}\,\,, \no \\
\{ \tilde{x},\, \tilde{p}^{\theta\rho\sigma} \}_{\phi =0} &=& \delta^{\rho\sigma}_{\mu\nu}\,\,, \no \\
\{ \tilde{x},\, \tilde{p}_\theta^{\rho\sigma} \}_{\phi =0} &=& - \{ \tilde{\pi}^\mu , \tilde{p}^{\rho\sigma}_\theta \} \,=\, \frac{1}{\theta} \eta^{\mu[\rho} v ^{\sigma]} \,-\,\frac{4}{\theta^4} (\theta v)^\mu \, \theta^{\rho\sigma}\,\,.
\eaa

The new Hamiltonian $\tilde{H}$ is given by $\tilde{H}=H_0 (\tilde{q}, \tilde{p}, \tilde{\lambda} )$ where $H$ was given by \cite{alexei0} and $\tilde{\lambda}$ is the modified Lagrangian multipliers for the constraints.   So, we have that

\bee
\label{}
\tilde{H}\,=\,\frac{\tilde{e}}{2} \Big( \tilde{v}^2 \,-\,m^2 \Big)\,+\, \tilde{\lambda}_{1\mu} \tilde{G}^\mu\,+\, \tilde{\lambda}_{2\mu} \tilde{T}^\mu \,+\,\tilde{\lambda}_e \tilde{p}_e\,+\, \tilde{\lambda}_{\theta \mu\nu}\,\tilde{p}^{\mu\nu}_\theta\,\,,
\eee

\ni which is a general expression where $\tilde{\lambda}_{2\mu} = 0$ since the standard $\lambda^{\mu}_2 = 0$.   And we have that

\bee
\label{}
\tilde{\lambda}_1^\mu \,=\,\tilde{e}\tilde{v}^\mu \,+\,\frac{2}{\theta^2} \Big(\tilde{\lambda}_\theta \tilde{v} \Big)^\mu \,-\, \frac{4}{\theta^4} \Big( \tilde{\theta}\tilde{\lambda}_\theta \Big) \Big( \tilde{\theta}\tilde{v} \Big)^\mu\,\,.
\eee

\ni The next step of BFFT aims at the constraints, so the extended constraints can be given by

\bee
\label{A}
\tilde{G}^\mu\,=\, p^\mu \,-\,v^\mu\,+\,\omega^\mu
\eee

\ni and

\bee
\label{B}
\tilde{T}_\mu \,=\,\pi_\mu\,+\,p_\mu^\omega \,+\, p_\mu^\xi \,+\, \frac{\theta_{\mu\nu}}{\theta^2}\, W^\nu\,\,,
\eee

\ni where $W^{\nu}$ was given above and from \eqref{A} and \eqref{B} we can write that

\baa
\label{C}
&&\{ \tilde{T}, \tilde{T} \} \,=\,\{ \tilde{G}, \tilde{G} \} \,=\,\{ \tilde{G}, \tilde{T} \} \,=\,0 \,\,,\no \\
&&\{ \tilde{G}, \tilde{p_e} \} \,=\,\{ \tilde{T}, \tilde{p}_\theta^{\rho\sigma} \} \,=\,\{ \tilde{T}, \tilde{p}_e \} \,=\,\{ \tilde{G}, \tilde{p}_\theta^{\rho\sigma} \} \,=\,0\,\,, \no \\
&&\{ \tilde{p}_\theta^{\alpha\beta}, \tilde{p}_\theta^{\rho\sigma} \} \,=\,\{ \tilde{p}_\theta^{\alpha\beta}, \tilde{p}_e \} \,=\,\{ \tilde{p}_e, \tilde{p}_e \} \,=\,0
\eaa

\ni and obviously we have all the constraints as being first-class ones.   The first-class Lagrangian is given by

\bee
\label{D}
\tilde{\cal L} \,=\, \dot{\tilde{x}}^\mu\,\tilde{v}_\mu \,-\,\frac{\tilde{e}}{2} \Big( \tilde{v}^2 \,-\, m^2 \Big) 
\,+\, \frac{1}{\tilde{\theta}^2} \dot{\tilde{v}}_\mu \,\tilde{\theta}^{\mu\nu} \,\tilde{v}_\nu
\eee

\ni which, using \eqref{7.2}, can be demonstrated explicitly to be gauge invariant.  And again, if we make the auxiliary variables equal to zero we obtain the initial Lagrangian.

We can see that this last Lagrangian is totally different from the one we have obtained in Eq. \eqref{6.21}.  In both methods we have introduced auxiliary variables to reach gauge invariance.  However, in this second result, the auxiliary variables are part of the phase space, and in \eqref{6.21} the initial phase space is maintained. But, in \eqref{D} the number of degrees of freedom is conserved since the relation between variables and constraints is kept.  On the other hand, in  \eqref{6.21} we have a way more complicated form of the Lagrangian.  In conclusion we can consider that both methods demonstrate that, besides that it is demonstrated that gauge invariance of the original Lagrangian can be analytically obtained, there is an entire family of gauge transformations.


\section{Considerations and final remarks}

During the last seventeen years, since the work of Seiberg and Witten \cite{sw}, besides the interest in string theories, we have been witnessing a growing interest in NC space-time theories.  Both problems concerning quantum mechanics and field theories described in NC space have been explored in an excited way together with the hope that noncommutativity would be the natural path to unify quantum mechanics and gravitation.   This last one is one of the targets of string theories.  The connection between string theories and NC space theories was established in the above paper \cite{sw}, which can lead us to agree with the existence of a NC gravity.

In this paper we have considered one aspect of the noncommutativity of position coordinates which, besides the mentioned string algebra, can be obtained also as a result of a canonical quantization of dynamical models where the position coordinates play the role of operators, which lead us to a NC quantum mechanics and consequently to a NC quantum theory.

Having said that, in this work we have analyzed the NC version of the relativistic particle and its gauge invariance issue.   Since it is a demonstrated second-class system \cite{alexei0}, the obtainment of gauge invariant formulation of this model is an interesting subject since gauge invariance is a fundamental issue in theoretical physics and specifically one of the foundations of the standard model, these are our main motivations.

In this way we have worked with two different starting points and two different paths to consider this task.  Our objective is to make a comparison and to establish if the gauge invariant action is unique or not, analytically speaking.  To begin from two different starting points is also useful to generalize the approach.

Firstly we have attacked the action itself, using the Noether procedure to obtain a very specific gauge invariant action.  After that we have analyzed its constraint algebra and we converted this second-class algebra into a first-class one through the well known BFFT method, which was never applied to NC models in the literature.  In both approaches we have to introduce auxiliary variables. In both methods, as it is expected, the number of degrees of freedom is conserved.

Since we have succeed in both tasks we can conclude that the NC action for the relativistic particle has a family of gauge invariant actions.  A convenient way to disclose the whole family is through the well known Faddeev-Jackiw method which, via the fact that its so-called zero mode can be conveniently chosen in order to reveal this mentioned group of gauge transformations.  It is an ongoing research which will be published elsewhere.

\section{Acknowledgments}

\ni The authors thank CNPq (Conselho Nacional de Desenvolvimento Cient\' ifico e Tecnol\'ogico), Brazilian scientific support federal agency, for partial financial support, Grants numbers 302155/2015-5, 302156/2015-1 and 442369/2014-0 and E.M.C.A. thanks the hospitality of Theoretical Physics Department at Federal University of Rio de Janeiro (UFRJ), where part of this work was carried out.


\end{document}